# CODE COVERAGE BASED TEST CASE SELECTION AND PRIORITIZATION

R.Beena [1], Dr.S.Sarala [2]

[1]Research Scholar, Dept. of Information Technology, Bharathiar University, Coimbatore.
[2]Assistant Professor, Dept. of Information Technology, Bharathiar University, Coimbatore

## ABSTRACT

*Regression Testing is exclusively executed to guarantee the desirable functionality of existing software after pursuing quite a few amendments or variations in it. Perhaps, it testifies the quality of the modified software by concealing the regressions or software bugs in both functional and non-functional applications of the system. In fact, the maintenance of test suite is enormous as it necessitates a big investment of time and money on test cases on a large scale. So, minimizing the test suite becomes the indispensable requisite to lessen the budget on regression testing. Precisely, this research paper aspires to present an innovative approach for the effective selection and prioritization of test cases which in return may procure a maximum code average.*

## 1. INTRODUCTION

Regression testing is an authentication method pursued in all levels of system and software testing. Despite ensuring the functioning capacity of the software or system after making amendments, Regression Testing, exhibits a predominant function with the previously deployed test codes of the enhanced software. The prime aspiration of running a Regression Test is to assure that modified or amended component of software does not give way for bugs in the unaltered portion of the software. The re-execution of test cases are performed to verify that the previous functionality clubbed with the present changes is desirably functioning.

The various regression testing techniques are test case minimization, test case selection and test case prioritization .The aim of test case minimization technique is to eliminate the redundant test cases, while test case selection techniques are performed to reduce the size of a test suite. Test case prioritization techniques are concerned with ordering of test cases for detection of faults at the earliest. This paper presents a customized technique for Test case selection and Test case prioritization.

Test case selection implies identifying a smaller subset of test suite from the existing large test suite [1]. According to [2], test case selection problem is stated subsequently.

> **Given**: The original program, P, the revised version of P, P' and a test suite, T.
> **Aim**: To identify T' ⊆ T, for the modified version P'

Test Case Prioritization is the process of arranging test cases in an order according to some criteria. Test case prioritization problem defined by Rothermel et al. [3] is follows:

> **Given**: A test suite, T, the set of permutations of T, PT, a function from PT to the real numbers, f.
> **Aim:** To identify T′ ∈ PT such that (∀ T″) (T″ ∈ PT) (T″ ≠ T′) f (T′) ≥ f (T″)

Here, 'PT' represents the set of all possible prioritizations of 'T' and 'f'. The function that is applied to any such ordering actually yields an award value for the respective ordering.

## 2. RELATED WORK

Fischer et al. formulated a test case selection problem with the application of Integer Programming [4]. The variations of the control flow were not discussed in this approach.

Agrawal et al. outlined an exclusive strategy on test case selection with a special perspective to the discrepancies found in the program slicing techniques [5].

Rothermel and Harrold elucidated regression test case selection techniques based on graph walking of Control, Program Dependence Graphs [6], and System Dependence Graphs [7] besides, Control Flow Graphs [8].

Benedusi et al. executed path analysis for test case selection [9]. A testing structure called TestTube was introduced by Chen et al. [10] which make use of a modification-based method for selection of test cases.

Leung and White highlighted a firewall technique for regression testing of system integration [11]. Laski and Szemer offered a technique for test case selection which is based on cluster identification technique [12].

In [13], [14], Rothermel et al. were the premiers to study test case prioritization predicaments that paved a way to them to present six different strategies based on the coverage of statement or branches.

In [15], Li et al. gives empirical study results of two metaheuristic search techniques and three greedy search techniques applied to six programs for regression test case prioritization.

In [16], Praveen et al. initiated a novel test case prioritization algorithm that calculates average faults observed per minute.

A Regression Testing Technique for Test Case Prioritization based on Code Coverage criteria is recommended by K.K.Aggarwal in [17].

## 3. TEST CASE SELECTION

The test cases those are available for the existing version of the program is grouped into three clusters. Those clusters are named as out-dated, required and surplus. The out-dated cluster contains the test cases that are not required by both the original program and the modified program. The required test case group consists of the test cases that are required to be executed for the modified version of the software. The surplus group comprises of test cases that may be essential for the later versions of P but are not required for the modified version of P i.e. P'. The algorithm for Test Case Selection (TCS) which is contributed in the previous work [18] is given in Figure 1.

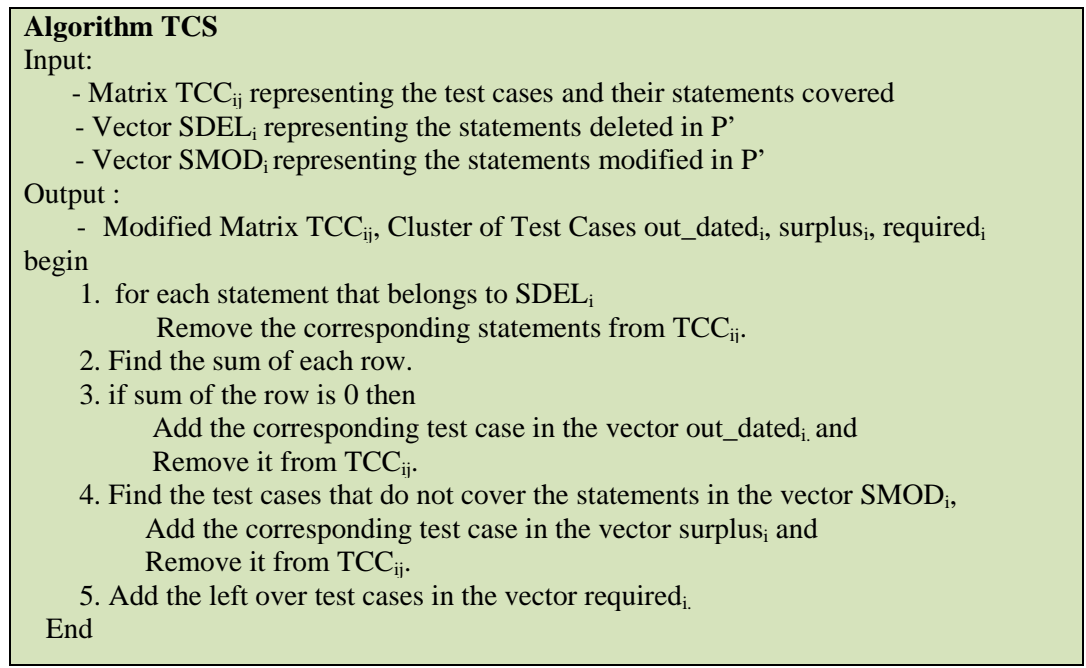

**Algorithm TCS**
Input:
 - Matrix $TCC_{ij}$ representing the test cases and their statements covered
 - Vector $SDEL_i$ representing the statements deleted in P'
 - Vector $SMOD_i$ representing the statements modified in P'
Output :
 - Modified Matrix $TCC_{ij}$, Cluster of Test Cases $out\_dated_i$, $surplus_i$, $required_i$
begin
 1. for each statement that belongs to $SDEL_i$
   Remove the corresponding statements from $TCC_{ij}$.
 2. Find the sum of each row.
 3. if sum of the row is 0 then
   Add the corresponding test case in the vector $out\_dated_i$ and
   Remove it from $TCC_{ij}$.
 4. Find the test cases that do not cover the statements in the vector $SMOD_i$,
   Add the corresponding test case in the vector $surplus_i$ and
   Remove it from $TCC_{ij}$.
 5. Add the left over test cases in the vector $required_i$.
End

Figure 1. Algorithm TCS

## 4. TEST CASE PRIORITIZATION

The output obtained from algorithm TCS is supplied as input to the algorithm Test Case Prioritization (TCP) which is described in Figure 2. An example for the steps of the algorithm TCS and TCP is elucidated in section 5.

```
Algorithm TCP
Input:
    - Modified Matrix TCC_ij representing selected test cases and their statements covered
Output :
    - Vector TCP_i which contains the test cases to achieve 100% code coverage.
begin
    1. Find the sum of each row of the matrix TCC_ij.
    2. Select the test case with highest sum and add that test case into the vector TCP_i.
    3. Remove all the statements covered by that test case.
    4.  Repeat step1 until all the statements are deleted.
End
```

Figure 2.  Algorithm TCP

## 5. EXPERIMENTS AND RESULTS

### 5.1. Test Case Selection

The original version of the program contains 15 statements and 15 test cases. The test cases and the coverage of the statements by the test cases are represented as a binary matrix. The binary matrix represented as ($TCC_{ij}$) is given in Table 1.

Table1.  Test cases and statement coverage $TCC_{ij}$

|     | S1 | S2 | S3 | S4 | S5 | S6 | S7 | S8 | S9 | S10 | S11 | S12 | S13 | S14 | S15 |
|-----|----|----|----|----|----|----|----|----|----|-----|-----|-----|-----|-----|-----|
| T1  | 1  | 0  | 1  | 1  | 1  | 0  | 1  | 1  | 0  | 1   | 1   | 0   | 0   | 0   | 0   |
| T2  | 1  | 0  | 0  | 1  | 0  | 1  | 0  | 1  | 1  | 1   | 0   | 0   | 1   | 0   | 0   |
| T3  | 0  | 0  | 1  | 0  | 0  | 1  | 0  | 0  | 0  | 1   | 0   | 0   | 1   | 0   | 0   |
| T4  | 0  | 1  | 0  | 1  | 0  | 0  | 1  | 1  | 1  | 0   | 0   | 0   | 1   | 1   | 0   |
| T5  | 1  | 0  | 1  | 0  | 1  | 1  | 0  | 0  | 0  | 1   | 0   | 0   | 0   | 1   | 1   |
| T6  | 0  | 1  | 0  | 1  | 0  | 0  | 1  | 0  | 1  | 1   | 1   | 1   | 0   | 0   | 0   |
| T7  | 1  | 0  | 0  | 0  | 1  | 0  | 0  | 0  | 1  | 0   | 0   | 1   | 0   | 1   | 0   |
| T8  | 0  | 1  | 1  | 0  | 0  | 1  | 0  | 0  | 1  | 0   | 1   | 1   | 0   | 0   | 0   |
| T9  | 0  | 0  | 0  | 1  | 1  | 1  | 1  | 0  | 1  | 1   | 0   | 0   | 1   | 0   | 0   |
| T10 | 1  | 1  | 0  | 0  | 0  | 0  | 1  | 1  | 1  | 0   | 0   | 0   | 1   | 1   | 1   |
| T11 | 0  | 0  | 0  | 1  | 0  | 0  | 0  | 1  | 0  | 1   | 0   | 0   | 1   | 0   | 0   |
| T12 | 1  | 1  | 0  | 0  | 1  | 0  | 1  | 0  | 0  | 0   | 0   | 0   | 0   | 0   | 1   |
| T13 | 0  | 1  | 0  | 1  | 0  | 1  | 1  | 1  | 0  | 1   | 0   | 1   | 1   | 1   | 0   |
| T14 | 0  | 1  | 0  | 0  | 0  | 0  | 0  | 0  | 0  | 0   | 0   | 0   | 0   | 0   | 0   |
| T15 | 1  | 0  | 0  | 0  | 1  | 0  | 0  | 0  | 1  | 0   | 1   | 1   | 0   | 0   | 0   |

Let us consider, in the modified version of the program the statements S3, S4, S6, S8, S10, S13 have been deleted and the statements S2, S7, S15 have been modified. So the two vectors $SDEL_i$ and $SMOD_i$ are represented as

$SDEL_i = \{S3, S4, S6, S8, S10, S13\}$   $SMOD_i = \{S2, S7, S15\}$

Table2. Modified $TCC_{ij}$

|  | S1 | S2 | S5 | S7 | S9 | S11 | S12 | S14 | S15 |
|---|---|---|---|---|---|---|---|---|---|
| **T1** | 1 | 0 | 1 | 1 | 0 | 1 | 0 | 0 | 0 |
| **T2** | 1 | 0 | 0 | 0 | 1 | 0 | 0 | 0 | 0 |
| **T3** | 0 | 0 | 0 | 0 | 0 | 0 | 0 | 0 | 0 |
| **T4** | 0 | 1 | 0 | 1 | 1 | 0 | 0 | 1 | 0 |
| **T5** | 1 | 0 | 1 | 0 | 0 | 0 | 0 | 1 | 1 |
| **T6** | 0 | 1 | 0 | 1 | 1 | 1 | 1 | 0 | 0 |
| **T7** | 1 | 0 | 1 | 0 | 1 | 0 | 1 | 1 | 0 |
| **T8** | 0 | 1 | 0 | 0 | 1 | 1 | 1 | 0 | 0 |
| **T9** | 0 | 0 | 1 | 1 | 1 | 0 | 0 | 0 | 0 |
| **T10** | 1 | 1 | 0 | 1 | 1 | 0 | 0 | 1 | 1 |
| **T11** | 0 | 0 | 0 | 0 | 0 | 0 | 0 | 0 | 0 |
| **T12** | 1 | 1 | 1 | 1 | 0 | 0 | 0 | 0 | 1 |
| **T13** | 0 | 1 | 0 | 1 | 0 | 0 | 1 | 1 | 0 |
| **T14** | 0 | 1 | 0 | 0 | 0 | 0 | 0 | 0 | 0 |
| **T15** | 1 | 0 | 1 | 0 | 1 | 1 | 1 | 0 | 0 |

The matrix $TCC_{ij}$ is personalized by removing the statements that are available in the vector group $SDEL_i$ at the end of the execution of step 1.The modified $TCC_{ij}$ is given in Table 2. The number of statements covered by each test case is calculated according to step 2. For example T1 covers four statements namely S1, S5, S7 and S11.Table 3 represents the total number of statements covered by each test case.

Table3. Number of statements covered by test cases

| Test Cases | Statements Covered |
|---|---|
| T1 | 4 |
| T2 | 2 |
| T3 | 0 |
| T4 | 4 |
| T5 | 4 |
| T6 | 5 |
| T7 | 5 |
| T8 | 4 |
| T9 | 3 |
| T10 | 6 |
| T11 | 0 |
| T12 | 5 |
| T13 | 4 |
| T14 | 1 |
| T15 | 5 |

As given in step 3, the test cases with the sum as zero are removed from the matrix $TCC_{ij}$. Now the new matrix $TCC_{ij}$ is given in Table 4. A new vector $out\_dated_i$ is created which contains the removed test cases from $TCC_{ij}$. The vector $out\text{-}dated_i = \{T3, T11\}$

Table4. $TCC_{ij}$ without out-dated$_i$

|     | S1 | S2 | S5 | S7 | S9 | S11 | S12 | S14 | S15 |
|-----|----|----|----|----|----|-----|-----|-----|-----|
| T1  | 1  | 0  | 1  | 1  | 0  | 1   | 0   | 0   | 0   |
| T2  | 1  | 0  | 0  | 0  | 1  | 0   | 0   | 0   | 0   |
| T4  | 0  | 1  | 0  | 1  | 1  | 0   | 0   | 1   | 0   |
| T5  | 1  | 0  | 1  | 0  | 0  | 0   | 0   | 1   | 1   |
| T6  | 0  | 1  | 0  | 1  | 1  | 1   | 1   | 0   | 0   |
| T7  | 1  | 0  | 1  | 0  | 1  | 0   | 1   | 1   | 0   |
| T8  | 0  | 1  | 0  | 0  | 1  | 1   | 1   | 0   | 0   |
| T9  | 0  | 0  | 1  | 1  | 1  | 0   | 0   | 0   | 0   |
| T10 | 1  | 1  | 0  | 1  | 1  | 0   | 0   | 1   | 1   |
| T12 | 1  | 1  | 1  | 1  | 0  | 0   | 0   | 0   | 1   |
| T13 | 0  | 1  | 0  | 1  | 0  | 0   | 1   | 1   | 0   |
| T14 | 0  | 1  | 0  | 0  | 0  | 0   | 0   | 0   | 0   |
| T15 | 1  | 0  | 1  | 0  | 1  | 1   | 1   | 0   | 0   |

The vector $SMOD_i$ contains the statements that are modified in the new version of the program and the test cases that do not cover those statements are removed from $TCC_{ij}$ and inserted into the cluster $surplus_i$. The new $TCC_{ij}$ is given in Table 5. The vector $surplus_i = \{T2, T7, T15\}$

Table5. $TCC_{ij}$ without surplus$_i$

|     | S1 | S2 | S5 | S7 | S9 | S11 | S12 | S14 | S15 |
|-----|----|----|----|----|----|-----|-----|-----|-----|
| T1  | 1  | 0  | 1  | 1  | 0  | 1   | 0   | 0   | 0   |
| T4  | 0  | 1  | 0  | 1  | 1  | 0   | 0   | 1   | 0   |
| T5  | 1  | 0  | 1  | 0  | 0  | 0   | 0   | 1   | 1   |
| T6  | 0  | 1  | 0  | 1  | 1  | 1   | 1   | 0   | 0   |
| T8  | 0  | 1  | 0  | 0  | 1  | 1   | 1   | 0   | 0   |
| T9  | 0  | 0  | 1  | 1  | 1  | 0   | 0   | 0   | 0   |
| T10 | 1  | 1  | 0  | 1  | 1  | 0   | 0   | 1   | 1   |
| T12 | 1  | 1  | 1  | 1  | 0  | 0   | 0   | 0   | 1   |
| T13 | 0  | 1  | 0  | 1  | 0  | 0   | 1   | 1   | 0   |
| T14 | 0  | 1  | 0  | 0  | 0  | 0   | 0   | 0   | 0   |

All the remaining test cases that are available in $TCC_{ij}$ are inserted into a new cluster group $required_i$ as mentioned in step 5. The vector

$required_i = \{T1, T4, T5, T6, T8, T9, T10, T12, T13, T14\}$

The comparison between the original size of the test suite and the reduced size of the test suite is specified in Figure 3. The result shows that there is a notable reduction in the size between the two test suites.

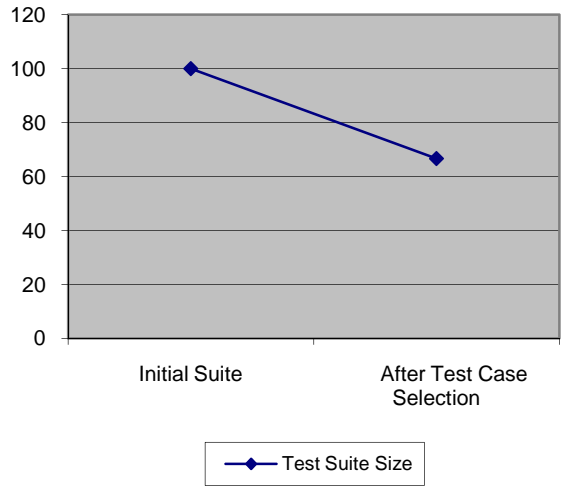

Figure3. Test Suite Size after Selection

## 5.2. Test Case Prioritization

Input matrix $TCC_{ij}$ for Test Case Prioritization is given in Table 5.

Iteration 1:

As given in step 1, the number of statements covered by each test case is counted from the new matrix $TCC_{ij}$. It is given in Table 6.

Table 6. Number of statements covered by test cases

| Test Cases | Statements Covered |
|---|---|
| T1 | 4 |
| T4 | 4 |
| T5 | 4 |
| T6 | 5 |
| T8 | 4 |
| T9 | 3 |
| T10 | 6 |
| T12 | 5 |
| T13 | 4 |
| T14 | 1 |

As given in step 2, the test case with highest sum is removed from $TCC_{ij}$ and that test case is added into the Test Case Prioritized vector $TCP_i$. The vector $TCP_i = \{T10\}$. All the statements that are covered by the test case T10 is removed from $TCC_{ij}$. The modified $TCC_{ij}$ is given in Table 7.

Table7.Updated TCC$_{ij}$

Iteration 2:

As given in step 1, the sum of each row of the updated matrix TCC$_{ij}$ given in Table 7 is computed and the sum is specified in Table 8.

Table8. Number of statements covered by test cases

| Test Cases | Statements Covered |
|---|---|
| T1 | 2 |
| T4 | 0 |
| T5 | 1 |
| T6 | 2 |
| T8 | 2 |
| T9 | 1 |
| T10 | 0 |
| T12 | 1 |
| T13 | 1 |
| T14 | 0 |

As given in step 2, the test case with highest sum is removed from TCC$_{ij}$ and that test case is added into the vector TCP$_i$. Here in this example, there are three test cases {T1, T6, T8} with highest sum. The test case T1 is selected here. The issue of equal priority is to be considered in future. Now the vector TCP$_i$ = {T10, T1}. All the statements that are covered by the test case T1 is removed from TCC$_{ij}$. The modified TCC$_{ij}$ is given in Table 9.

International Journal of Software Engineering & Applications (IJSEA), Vol.4, No.6, November 2013International Journal of Software Engineering & Applications (IJSEA), Vol.4, No.6, November 2013

Table 9. Updated $TCC_{ij}$

|  | S5 | S11 | S12 |
|---|---|---|---|
| T1 | 1 | 1 | 0 |
| T4 | 0 | 0 | 0 |
| T5 | 1 | 0 | 0 |
| T6 | 0 | 1 | 1 |
| T8 | 0 | 1 | 1 |
| T9 | 1 | 0 | 0 |
| T10 | 0 | 0 | 0 |
| T12 | 1 | 0 | 0 |
| T13 | 0 | 0 | 1 |
| T14 | 0 | 0 | 0 |

|  | S12 |
|---|---|
| T1 | 0 |
| T4 | 0 |
| T5 | 0 |
| T6 | 1 |
| T8 | 1 |
| T9 | 0 |
| T10 | 0 |
| T12 | 0 |
| T13 | 1 |
| T14 | 0 |

Iteration 3:

As given in step 1, the sum of each row of the updated matrix $TCC_{ij}$ given in Table 9 is computed and the sum is specified in Table 10.

Table 10. Number of statements covered by test cases

| Test Cases | Statements Covered |
|---|---|
| T1 | 0 |
| T4 | 0 |
| T5 | 0 |
| T6 | 1 |
| T8 | 1 |
| T9 | 0 |
| T10 | 0 |
| T12 | 0 |
| T13 | 1 |
| T14 | 0 |

As given in step 2, the test case with highest sum is removed from $TCC_{ij}$ and that test case is added into the vector $TCP_i$. Here in this example, there are three test cases {T6, T8, T13} with highest sum. The test case T6 is selected here. The final prioritized vector

$$TCP_i = \{T10, T1, T6\}$$

Figure4 gives the size of the test suite after test case prioritization. The size of the test suite is very much reduced and hence the cost of regression testing and time for execution of test cases can be minimized to a greater extent.

47

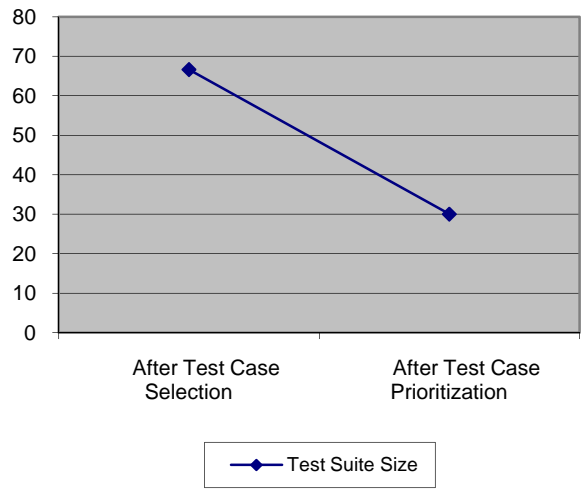

Figure4. Test Suite Size after Prioritization

## 6. CONCLUSION

Regression testing is carried out in the maintenance phase of the software development to retest the software for the revisions it has endured and to confirm the accurate functionalities of the revised version. A new technique for test case selection and test case prioritization process for regression testing is proposed in this paper. The proposed technique is very effective in terms of cost and time involved in regression testing. In future, the regression testing techniques may be combined with optimization algorithms to contribute more enhanced results.